\documentclass[twocolumn,showpacs,prc]{revtex4}
\usepackage{graphicx}
\usepackage{dcolumn}
\usepackage{bm}

\begin{document}

\title{Viscosity of High Energy Nuclear Fluids}

\author{V. Parihar}
\affiliation{Physics Department, Boston University, Boston MA 02215}
\author{ A.Widom, D. Drosdoff}
\affiliation{Physics Department, Northeastern University, Boston MA 02115}
\author{Y.N. Srivastava}
\affiliation{Physics Department \& INFN, University of Perugia, Perugia IT}

\begin{abstract}
Relativistic high energy heavy ion collision cross sections have been interpreted 
in terms of almost ideal liquid droplets of nuclear matter. The experimental low 
viscosity of these nuclear fluids have been of considerable recent quantum 
chromodynamic interest. The viscosity is here discussed in terms of the string 
fragmentation models wherein the temperature dependence of the nuclear fluid 
viscosity obeys the Vogel-Fulcher-Tammann law.    
\end{abstract}

\maketitle

\section{Introduction \label{IN}}

There has been considerable interest in probing experimentally 
a theoretical quark-gluon plasma phase transition by employing 
ultra-relativistic two heavy nuclei scattering. It is not presently 
clear whether or not a deconfined plasma has actually been observed. 
However, such scattering experiments have apparently produced nuclear 
matter at very high energy density wherein the matter acts as a 
relativistic almost {\em ideal} 
fluid\cite{Heinz:2005,Shuryak:2004a,Shuryak:2004b,Shuryak:2004c}; 
i.e. a quark-gluon matter fluid with very low 
viscosity\cite{Teaney:2003,Moore:2005,Casalderrey-Solana:2006,Gyulassy:2005}. 

In previous work\cite{Parihar:2006a,Parihar:2006b}, we argued from the 
string fragmentation model, that a high energy density fluid may exhibit 
a glass-like behavior with meson and baryon strings playing a role 
analogous to polymers in some condensed matter glasses. At very high 
temperatures, the viscosity of the stringy fluid is actually quite small. 
It is this regime (of an almost ideal fluid) that is of central importance 
in ultra-relativistic heavy ion scattering. Our purpose is to study how 
the viscosity of resulting nuclear matter varies with energy density, pressure 
and baryon density.  

In Sec.\ref{PF} we review the Kubo formula for the frequency dependent 
viscosity of a fluid and derive sum rules which allow for the determination 
of the zero frequency transverse viscosity in terms of the relaxation time for 
the non-diagonal elements of the pressure tensor. In Sec.\ref{QT}, we consider 
the quantized unit of circulation which contributes to turbulent eddy currents 
in the fluid. A bound for the Reynolds number of quantum turbulence leads to a 
quantum lower bound for the viscosity of the type that has been previously 
conjectured\cite{Shuryak:2005}. We review the low energy well established 
liquid drop model in Sec.\ref{OLDM}  and argue that the low energy density 
nuclear fluid is highly viscous. We turn to high energy nuclear matter in 
Sec.\ref{GCS} and in particular discuss the formal conformal symmetry associated 
with glue. This system is thought to describe high energy QCD nuclear matter in 
the limit of low baryon density. In Sec.\ref{lund} we review the notion of 
the QCD inspired string model and employ in Sec.\ref{VString} the string 
fragment ion picture to compute the viscosity of high temperature nuclear matter. 
The viscosity reaches small values in the ultra high temperature limit. 
In the concluding Sec.\ref{conc}, we compare the string fragmentation 
(jet) picture  with the quark gluon plasma view of high temperature nuclear matter.

\section{Pressure Fluctuations}
\label{PF} 

For an isotropic fluid in thermal equilibrium, the off diagonal elements of the pressure 
tensor (say \begin{math} P_{xy}({\bf r},t) \end{math}) fluctuations are described by the 
correlation function, 
\begin{equation}
K({\bf r}-{\bf r}^\prime ,t)=\frac{1}{\hbar }\int_0^\beta 
\left<P_{xy}({\bf r},t)P_{xy}({\bf r}^\prime ,-i\lambda)\right>d\lambda ,
\label{PF1}
\end{equation}   
wherein \begin{math} \beta =(\hbar/k_BT) \end{math}. The complex frequency dependent fluid 
shear viscosity  \begin{math} \eta (\zeta ) \end{math} is determined by the Kubo 
formalism\cite{wang:1996,wang:1999,Thoma:1991,Gupta:1997,Nakamura:2005}
\begin{eqnarray}
{\cal G}(t)=\int K({\bf r} ,t) d^3{\bf r},
\nonumber \\ 
\eta (\zeta )=\int_0^\infty {\cal G}(t) e^{i\zeta t} dt,
\ \ \ {\Im m}\ \zeta >0.
\label{PF2}
\end{eqnarray}
With \begin{math} \rho  \end{math} as the mass density of the fluid and 
\begin{math} c_\infty  \end{math} as the velocity of a finite high frequency transverse 
sound, one finds the dispersion relations 
\begin{eqnarray}
{\cal G}(t)=\frac{2}{\pi }\int_0^\infty 
{\Re e}\eta (\omega +i0^+) \cos(\omega t) d\omega , 
\nonumber \\ 
\eta (\zeta )=-\frac{2i\zeta }{\pi }\int_0^\infty 
\frac{{\Re e}\eta (\omega +i0^+)d\omega }{\omega^2-\zeta^2},
\label{PF3}
\end{eqnarray} 
together with the sum rule 
\begin{equation}
{\cal G}(0)\equiv \rho c_\infty ^2=\frac{2}{\pi }
\int_0^\infty {\Re e}\eta (\omega +i0^+) d\omega . 
\label{PF4}
\end{equation}
In the low frequency limit 
\begin{equation}
\eta \equiv \lim_{\omega \to 0}{\Re e}\eta (\omega +i0^+)
=\int_0^\infty {\cal G}(t)dt,
\label{PF5}
\end{equation}
the relaxation time \begin{math} \tau  \end{math} for the decay of the correlation function 
\begin{equation}
\tau =\frac{1}{{\cal G}(0)}\int_0^\infty {\cal G}(t)dt=\frac{1}{\rho c_\infty ^2}\ \eta 
\label{PF6}
\end{equation}
uniquely determines the shear viscosity 
via\cite{Danielewicz:1985,Baym:1990,Arnold:2000,Arnold:2003}
\begin{equation}
\eta =\rho c_\infty ^2 \tau . 
\label{PF7}
\end{equation}
To find the viscosity at zero frequency for a fluid of mass density 
\begin{math} \rho  \end{math} employing the Kubo formalism, one must determine 
the velocity of transverse sound \begin{math} c_\infty  \end{math} and the 
relaxation time \begin{math} \tau  \end{math} of the off diagonal elements of 
the pressure. 

\section{Quantum Turbulence}
\label{QT}

The following arguments, first applied by Feynman\cite{Feynman:1955} to superfluid Helium, 
hold true for any fluid whose constituent particles all have mass \begin{math} M \end{math}. 
For flows with eddy currents on a velocity scale \begin{math} V \end{math} and a 
length scale \begin{math} L \end{math}, the {\em circulation} of the eddy velocity field 
is given by 
\begin{equation}
\oint {\bf v}\cdot d{\bf r}=VL
\label{QT1}
\end{equation}
From quantum fluid mechanics, the Feynman number \begin{math} {\cal F} \end{math} 
of quantum vortices in a circulating eddy is given by the Bohr quantization of circulation 
\begin{equation}
M\oint {\bf v}\cdot d{\bf r}=\oint {\bf p}\cdot d{\bf r}=2\pi \hbar {\cal F};
\label{QT2}
\end{equation}
i.e. the Feynman number is 
\begin{equation}
{\cal F}=\frac{MVL}{2\pi \hbar}.
\label{QT3}
\end{equation}
On the other hand, the Reynolds number for a circulating eddy is well known to be 
\begin{equation}
{\cal R}=\frac{\rho VL}{\eta } ,
\label{QT4}
\end{equation} 
yielding a ratio which depends only on intrinsic fluid properties 
and not on the nature of the eddy flow; 
\begin{equation}
\frac{\cal F}{\cal R}=\frac{ M \eta }{2\pi \hbar \rho }.
\label{QT5}
\end{equation}
If \begin{math} \rho = Mn \end{math}, with \begin{math} n \end{math} as the number 
of constituent particles per unit volume, then 
\begin{equation}
\frac{\cal F}{\cal R}=\frac{\eta }{2\pi \hbar n}
\ \ \ ({\rm non-relativistic}).
\label{QT6}
\end{equation}
Since the Feynman number of quantized vortices in an eddy is larger than the 
Reynolds number of the eddy, we have the quantum viscosity inequality 
\begin{equation}
{\cal F}>{\cal R}\ \ \ \ \Rightarrow
\ \ \ \ \eta > 2\pi \hbar n.
\label{QT7}
\end{equation}
The above lower bound is found to apply to all known chemically pure substances. 
For example, for pure water 
\begin{math} {\cal F}\approx 45 {\cal R} \gg {\cal R} \end{math}.

\section{Old Liquid Drop Model}
\label{OLDM}

The first nuclear fluid model was employed in a non-relativistic context. In the 
liquid drop model, one has a momentum sphere of filled quasi-particle states out to 
a Fermi momentum \begin{math} p_F=\hbar k_F \end{math}. The number of nucleons 
\begin{math} A \end{math} in a spherical nucleus of radius \begin{math} R \end{math} 
can be computed by filling up a Fermi sphere in momentum space and a sphere in position 
space 
\begin{equation}
A=\frac{4}{(2\pi)^3}\left(\frac{4\pi k_F^3}{3}\right)
\left[\frac{4\pi R^3}{3}\right]=\frac{8(k_FR)^3}{9\pi }.
\label{OLDM1}
\end{equation}
The liquid drop nucleus is nearly incompressible so that the radius obeys
\begin{equation}
R=aA^{1/3}\ \ \ {\rm wherein}\ \ \ a\approx 1.2\times 10^{-13} {\rm cm}.
\label{OLDM2}
\end{equation}
The Fermi velocity for a nearly ideal Fermi fluid is given by  
\begin{eqnarray}
v_F=\frac{1}{\hbar}\frac{dE_F}{dk_F}
\approx \frac{\hbar k_F}{M}=\left(\frac{9\pi}{8}\right)^{1/3}\frac{\hbar }{Ma}\ .
\label{OLDM3}
\end{eqnarray}
Note that \begin{math} (v_F/c)\approx 0.27 \end{math} which implies that the nucleons 
within the droplet are not entirely non-relativistic. 

The density of energy states per unit volume at the Fermi surface is given by 
\begin{equation}
g_F=\frac{4}{(2\pi)^3}\left(4\pi k_F^2 \frac{dk_F}{dE_F}\right)
=\frac{2}{\pi^2}\left(\frac{k_F^2}{\hbar v_F}\right).
\label{OLDM4}
\end{equation}
The specific heat per unit volume of a Fermi liquid is given by 
\begin{equation}
c=T\frac{ds}{dT}=\left(\frac{\pi ^2 k_B^2 g_F}{3}\right)T+\ldots 
\ \ {\rm as}\ \ T\to 0.
\label{OLDM5}
\end{equation}
Which implies a low temperature entropy per unit volume 
\begin{equation}
s\approx \left(\frac{\pi ^2 k_B^2 g_F}{3}\right)T
=\left(\frac{\pi ^2 k_B^2 g_F}{3}\right)\frac{d\epsilon }{ds}
\label{OLDM6}
\end{equation}
wherein \begin{math} \epsilon \end{math} is the excitation energy per 
unit volume. For weakly excited states the entropy per unit volume obeys 
\begin{equation}
s=k_B\left(\frac{2\pi ^2 g_F}{3}\right)^{1/2}\sqrt{\epsilon }
=2k_Bk_F \sqrt{\frac{\epsilon}{3\hbar v_F}}\ .
\label{OLDM7}
\end{equation}
In terms of the total droplet entropy and energy, the thermodynamic equation of state 
for excitation energy \begin{math} E \end{math} weakly above the ground state energy 
\begin{math} E_{0A} \end{math} is given by   
\begin{equation}
\frac{S(E)}{k_B}=\frac{4}{3}A
\left(\frac{\pi k_F^2a^3}{\hbar v_F}\right)^{1/2}\sqrt{\frac{E-E_{0A}}{A}}\ .
\label{OLDM8}
\end{equation}
The density of excited energy levels for the droplet as a whole then reads 
\begin{equation}
D(E)\approx D(E_{0A})\exp\left[\frac{S(E)}{k_B}\right],
\label{OLDM9}
\end{equation}
which is experimentally reasonably accurate for heavy nuclei; a logarithmic 
plot of the density of excited states above the ground state yields a square 
root behavior\cite{Psukada:1966},
\begin{equation}
\ln\left[\frac{D(E)}{D(E_{0A})}\right]
\approx \sqrt{\frac{E-E_{0A}}{\tilde{E}_A}} 
\ \ {\rm as}\ \ E \to E_{0A}+0^+.
\label{OLDM10}
\end{equation}
In the opposite limit \begin{math} E\gg E_{0A} \end{math}, one finds  
\begin{math} S(E)\approx E/T_0  \end{math} wherein \begin{math} T_0 \end{math} 
is the Hagedorn temperature\cite{Hagedorn:1965}.

The viscosity of the Fermi fluid in the liquid drop is quite high. This may be 
understood by writing Eq.(\ref{PF7}) in the form 
\begin{equation}
\eta=v_F^2 \tau_c.
\label{OLDM11}
\end{equation}
wherein \begin{math} \tau_c \end{math} is an appropriate collision time for 
quasi-particle scattering. From two body scattering phase space considerations, 
one finds that \begin{math} \tau_c \propto T^{-2} \end{math} which implies a 
low nuclear energy liquid drop viscosity given by 
\begin{equation}
\eta\approx K\hbar 
\left(\frac{mv_F}{\hbar } \right)^5\left(\frac{\hbar v_F}{k_B T}\right)^2 
\label{OLDM12}
\end{equation}
in which \begin{math} K \end{math} is a dimensionless quantity of order unity,  
depending on the angular variation of the two body cross section. 
At very low temperatures, the viscosity grows very high. It will be 
discussed in what follows that as the temperature grows ever larger,  
the viscosity diminishes to smaller values until it is quantum limited 
and the fluid is nearly ideal. 

\section{Glue and Conformal Symmetry}
\label{GCS}
If \begin{math} G^a_{\mu \nu}  \end{math} denotes the gluon field, 
\begin{equation}
G^a_{\mu \nu }=\partial_\mu A^a_\nu -\partial_\nu A^a_\mu 
+\left(\frac{g}{\hbar c}\right)
f^a_{bc}A^b_\mu A^c_\nu ,
\label{GCS1}
\end{equation}
then the action describing pure glue reads 
\begin{equation}
W=-\frac{1}{2c}\int G^a_{\mu \nu }G_a^{\mu \nu } d^4x.
\label{GCS2}
\end{equation} 
Under the conformal scale changes in space time 
\begin{eqnarray}
x\to \bar{x}=\lambda x, 
\nonumber \\ 
A^a_\mu (x)\to \bar{A}^a_\mu (\bar{x})=\frac{1}{\lambda }A^a_\mu (x),
\nonumber \\ 
G^a_{\mu \nu }(x)\to \bar{G}^a_{\mu \nu }(\bar{x})=
\frac{1}{\lambda^2} G^a_{\mu \nu }(x),
\label{GCS3}
\end{eqnarray}  
the action is left invariant; 
i.e. \begin{math} W\to \bar{W}=W \end{math}. Such symmetry dictates that 
the trace of the energy-pressure tensor 
\begin{equation}
-\Theta=T^\mu_{\ \mu}=3P-u
\label{GCS4}
\end{equation}
is null. In Eq.(\ref{GCS4}), \begin{math} u \end{math} represents the 
energy per unit volume and \begin{math} P \end{math} represents the 
pressure. Formally, the energy-pressure tensor for glue is given by 
\begin{equation}
T^{Glue}_{\mu \nu }=\left(g^{\lambda \sigma }G^a_{\mu \sigma} G_{a\nu \lambda }
-\frac{1}{4}g_{\mu \nu }G^{a \mu \sigma}G_{a \mu \sigma} \right), 
\label{GCS5}
\end{equation}
which implies \begin{math}\Theta=0 \end{math} and which formally verifies 
the conformal invariance properties 
\begin{equation}
-\Theta^{Glue}=T^{\mu \ Glue}_{\ \ \mu }=0. 
\label{GCS6}
\end{equation}
On the other hand, for a theory with massive quarks the trace of the 
energy-pressure tensor is  
\begin{equation}
\Theta^{Quark}=\sum_f m_fc^2\bar{q}_fq_f. 
\label{GCS7}
\end{equation}
In the presence of Gluons, the trace of the energy-pressure tensor can reassert 
its strength via 
\begin{equation}
\Theta=\left<\sum_k m_kc^2\bar{q}_kq_k\right>. 
\label{GCS8}
\end{equation}
For example, for a heavy quark, say \begin{math} Q \end{math} with mass 
\begin{math} M \end{math}, in the presence of a low energy 
gluon field configuration one finds the trace anomaly\cite{Grundberg:1995} 
\begin{equation}
\Theta_Q=
Mc^2\left<0;G\left|\bar{Q}Q\right|0;G\right>
=-\left(\frac{2\alpha_s}{3\pi }\right)G^a_{\mu \nu}G_a^{\mu \nu},
\label{GCS9}
\end{equation}
where \begin{math} \alpha_s=g^2/4\pi \hbar c \end{math} is the QCD strong 
coupling strength. Employing purely thermodynamic reasoning with the 
Gibbs-Duhem equation, 
\begin{equation}
dP=sdT+\sum_f  n_f d\mu_f,
\label{GCS10}
\end{equation}
and the Euler relation for the energy density 
\begin{equation}
u=Ts-P+\sum_f \mu_fn_f,
\label{GC11}
\end{equation}
one finds the trace of the energy-pressure tensor 
\begin{eqnarray}
\Theta \equiv u-3P,
\nonumber \\ 
\Theta =Ts-4P+\sum_f \mu_f n_f, 
\nonumber \\ 
\Theta =T\left(\frac{\partial P}{\partial T}\right)_{\mu }-4P +\sum_f \mu_f 
\left(\frac{\partial P}{\partial \mu_f }\right)_{T,\mu_{k\ne f}}.
\label{GCS12}
\end{eqnarray}
If one replaces the quark chemical potentials with the quark activities 
\begin{equation}
z_f\equiv \exp(\mu_f/k_BT),
\label{GCS13}
\end{equation}
then the thermodynamic identity for the trace of the energy-pressure tensor 
has the more simple form 
\begin{equation}
\Theta =T\left(\frac{\partial P}{\partial T}\right)_z-4P,
\label{GCS14}
\end{equation} 
or equivalently, the energy density is related to the pressure according to 
\begin{equation}
u=T\left(\frac{\partial P}{\partial T}\right)_z-P.
\label{GCS15}
\end{equation} 

\begin{figure}[tp]
\scalebox {0.45}{\includegraphics{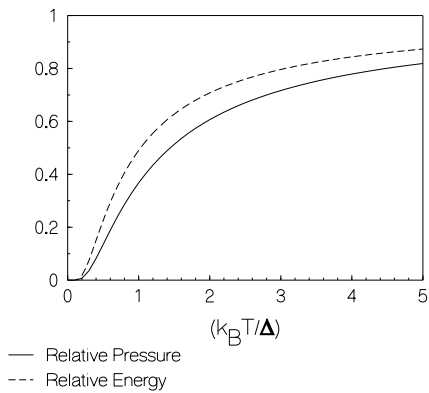}}
\caption{Shown are the relative pressure $P/P_\infty $ and relative energy 
$u/u_\infty $ plotted as a function of temperature.}
\label{fig1}
\end{figure}

Consider the following simple model of a proposed quark-gluon plasma 
deconfinement. At very high temperatures, one considers the quarks 
and gluons to act as a gas exhibiting conformal symmetry. 
This implies a pressure \begin{math} P_\infty  \end{math} and an energy 
density \begin{math} u_\infty  \end{math} obeying 
\begin{equation}
u_\infty=3P_\infty={\rm const.}\left[\frac{(k_BT)^4}{(\hbar c)^3}\right].
\label{GCS16}
\end{equation}
At low temperatures this ideal behavior cannot persist and the conformal 
symmetry is broken by an energy gap \begin{math} \Delta  \end{math}. 
The pressure may then be suppressed by a Boltzmann factor,
\begin{equation}
P=P_\infty e^{-\Delta /k_BT},
\label{GCS17}
\end{equation} 
which, in virtue of Eq.(\ref{GCS15}), implies 
\begin{equation}
u=u_\infty e^{-\Delta /k_BT}\left[1+\frac{\Delta }{3k_BT} \right].
\label{GCS18}
\end{equation} 
Plots of the quark-gluon deconfinement based on this model are shown in 
FIG. \ref{fig1}.
This simple model of the crossover between the confined phase and the  
quark-gluon plasma phase is in qualitative agreement with lattice QCD 
computations\cite{Karsch:2004}. The theoretical equilibrium statistical 
thermodynamic arguments presented here are not directly applicable to 
ultra high energy heavy ion nuclear collision for two reasons: 
(i) Very energetic collisions involve non-equilibrium processes and 
(ii) the collision fragments are most often described by QCD string 
fragmentation models\cite{Balakrishna:1993}.

\section{Lund String Model}
\label{lund}

\begin{figure}[bp]
\scalebox {0.50}{\includegraphics{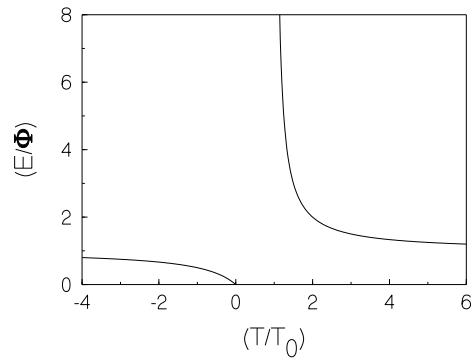}}
\caption{Shown is the energy of a single Lund string as a function of 
temperature. Note that the heat capacity of the string 
$C=(dE/dT)<0$ in the physical but metastable temperature ranges 
$-\infty < T < 0$ and $T_0 < T < \infty $.}
\label{fig2}
\end{figure}

To compute non-equilibrium processes wherein quarks are connected to one 
another via strings (gluon electric flux tubes) we begin with the activation 
entropy \begin{math} S_1(E) \end{math} of a single string; It is\cite{Parihar:2006a} 
\begin{eqnarray}
\frac{S_1}{k_B} = \left(\frac{E}{k_BT_0}\right)-
\frac{7}{2}\ln\left[\frac{E}{k_BT_0}\right]+\frac{\tilde{S}}{k_B}\ ,
\nonumber \\ 
\frac{\tilde{S}}{k_B} = \frac{1}{2}\left[\ln(3)+
7\ln\left(\frac{2\pi }{3}\right)\right], 
\label{lund1}
\end{eqnarray}
wherein the Hagedorn temperature \begin{math} T_0 \end{math} is related 
to the string tension \begin{math} \sigma \end{math} via 
\begin{equation}
k_BT_0=\sqrt{\frac{3\hbar c\sigma }{4\pi }} 
\approx 207\ {\rm MeV}.
\label{lund2}
\end{equation}
The numerical value of \begin{math} \sigma \end{math} (and thereby 
\begin{math} T_0 \end{math}) is found from the experimental slope of 
the lowest meson Regge trajectory. The energy of a single string as 
a function of temperature may be computed via the microcanonical 
statistical thermodynamic law, 
\begin{equation}
\frac{1}{T}=\frac{dS_1(E)}{dE},
\label{lund3}
\end{equation}
yielding,  
\begin{equation}
E=\Phi \left[\frac{T}{T-T_0}\right]
\ \ {\rm where}\ \  
\Phi=\frac{7k_BT_0}{2}\approx 725\ {\rm MeV}.
\label{lund4}
\end{equation}
The plot of energy as a function of temperature is given in 
FIG. \ref{fig2}. Note that the energy always decreases as the 
temperature increases. 
The heat capacity is negative in the physically allowed string temperature range; i.e.   
\begin{eqnarray}
C=\frac{dE}{dT}=-\frac{\Phi}{T_0}\left[\frac{T_0}{T-T_0}\right]^2<0, 
\nonumber \\ 
0 > T > = -\infty\ \ {\rm and}\ \ \infty > T > T_0.  
\label{lund5}
\end{eqnarray}
The physical interpretation of Eqs.(\ref{lund5}) is worthy of note. 
Since in accordance with the second law of thermodynamics, stable systems have a 
positive heat capacity we thereby deduce that a single string {\em during a collision} 
is in a metastable state. 

As discussed in previous work\cite{Parihar:2006a,Parihar:2006b}, the string near the 
beginning of the collision has a very low energy \begin{math} 0 < E < \Phi \end{math}.
Thus, the initial string temperature is negative. The string {\em cannot} have negative 
energy which implies that string temperatures between zero and the Hagedorn temperature 
are not allowed; i.e. \begin{math}0<T<T_0 \end{math} is unphysical.
As the energy of the string energy increases due to the absorption of 
collision energy \begin{math} E\to \Phi -0^+ \end{math}, the temperature decreases 
\begin{math} T \to -\infty \end{math}. When only a little bit more collision energy 
is absorbed by the string \begin{math} E=\Phi + 0^+ \end{math} the string temperature 
is virtually a positive infinite value. Any further cooling of the temperature requires 
more and more addition of the collision energy to be absorbed by a single string. 
We note in passing, that the notion of a negative temperature going positive by first 
approaching \begin{math} -\infty \end{math} and reappearing at \begin{math} +\infty \end{math} 
is {\em not} new. This property experimentally appears in laboratory nuclear magnetic 
resonant spin systems and laser media with inverted energy populations. 

For energy slightly above \begin{math} \Phi \end{math}, the string temperature is virtually  
infinite; i.e. \begin{math} T \to +\infty \end{math} as \begin{math} E\to \Phi +0^+ \end{math}. 
The question then arises (in the ultra high collision energy per nucleon limit) as to how much 
collision energy can be transferred to a single string in order to bring the temperature 
down from its very high temperature state. As will be made clear in the next Sec.\ref{VString}, 
the cooling time exponentially grows larger with decreasing temperature. Furthermore, the higher 
the collision energy per nucleon, the shorter the duration of a collision. When the duration 
of the collision is comparable to the cooling time for a given hot string temperature, the 
temperature stops getting smaller. The string temperatures stay high. While it is certainly 
possible to create many new strings with the ultra high collision energy, e.g. mesons, 
it will not be easy to cool the strings down from having a very high temperature. Let us 
now be more explicit regarding the time-scales in a stringy liquid droplet; i.e. the 
{\em new} high energy liquid droplet ``pancake'' model.

\section{Low Viscosity and the Ideal Fluid}
\label{VString}

The role of the entropy in determining transition rates follows from the 
the calculation rules of quantum mechanical transition rates; In particular one   
averages over initial states and sums over final states. In general the number of 
quantum states \begin{math} \Omega \end{math} is related to the entropy 
\begin{math} S \end{math} by the Boltzman formula 
\begin{equation}
S=k_B\ln \Omega .
\label{VString1}
\end{equation}
The ratio of the number of final states to the number of initial states is then 
determined by the exponential of an entropy difference 
\begin{equation}
\frac{\Omega_f}{\Omega_i}=\exp\left[\frac{S_f-S_i}{k_B}\right].
\label{VString2}
\end{equation}
Transition rates may be phase space dominated by the exponential 
entropy factors. The dominant entropy arises from bosonic gluon excitations 
of the QCD string. The Fermi or Bose nature of physical particles depend only 
on the number, respectively odd or even, of quarks tied together (in a polymer-like fashion) 
by the gluon color electric flux string. The thermal relaxation times for very hot 
\begin{math} T \to +\infty \end{math} strings to cool down to finite temperatures 
strings depends on the exponential of an entropy difference    
\begin{equation}
\frac{\tau }{\tau_\infty}=\exp\left[
\frac{S_1-S_{1 \infty}}{k_B}
\right].
\label{VString3}
\end{equation}
From Eqs.(\ref{lund1}) and (\ref{lund3}) one finds 
\begin{equation}
\frac{S_1}{k_B} = \frac{7}{2}\left(\frac{T_0}{T-T_0}\right)-
\left(\frac{7}{2}\right)\ln\left[\frac{T}{(T-T_0)}\right]+
\frac{S_{1 \infty}}{k_B}\ , 
\label{VString4}
\end{equation}
wherein
\begin{equation}
\frac{S_{1 \infty}}{k_B}=\frac{7}{2}\left[1+\ln\left(\frac{2\pi }{3}\right)
-\ln\left(\frac{7}{2}\right)\right]+\frac{1}{2}\ln 3. 
\label{VString5}
\end{equation}
Thus the cooling time scale from Eqs.(\ref{VString3}) and (\ref{VString4}) may 
be written as 
\begin{equation}
\tau = \tau_\infty \left(\frac{T-T_0}{T} \right)^{7/2}
\exp\left[{\Phi \over k_B(T-T_0)}\right].
\label{VString6}
\end{equation}

The viscosity implicit in QCD inspired string models is of a form which follows from 
Eqs.(\ref{PF7}) and (\ref{VString6}); It is 
\begin{equation}
\eta =\rho c_\infty^2 \tau_\infty 
\left(\frac{T-T_0}{T} \right)^{7/2}
\exp\left[{\Phi \over k_B(T-T_0)}\right].
\label{VString7}
\end{equation}
To compute the attempt frequency \begin{math} \tau_\infty ^{-1}  \end{math}
for a string to be created, for example between a quark anti quark pair, we need to 
consider the rate with which a quark in the ``negative vacuum Dirac-Fermi sea'' 
is excited into a positive energy state leaving behind a quark hole. The phase space 
element along the axis of the string connecting the quark and the hole is  
\begin{equation}
d^2\tilde{N}=\frac{dpdx}{2\pi \hbar}=\frac{dp}{dt}\left(\frac{dxdt}{2\pi \hbar}\right), 
\label{VString8}
\end{equation}
wherein the rate of change of momentum of the quark is the force, i.e. the string 
tension \begin{math} \sigma  \end{math}. Thus, the attempt frequency per unit time per 
unit length to excite a quark anti quark string, i.e. meson, is 
\begin{equation}
\frac{d^2\tilde{N}}{dtdx}=\frac{\sigma }{2\pi \hbar }. 
\label{VString9}
\end{equation}
For string of length \begin{math} L \end{math}, the energy is 
\begin{math} E=\sigma L \end{math} yielding the attempt frequency
\begin{equation}
\frac{1}{\tau_\infty }=\frac{d\tilde{N}}{dt}=\frac{E}{2\pi \hbar}=
\frac{\Phi }{2\pi \hbar}\left(\frac{T}{T-T_0}\right), 
\label{VString10}
\end{equation}
wherein Eq.(\ref{lund4}) has been invoked. 

From Eqs.(\ref{VString7}) and (\ref{VString10}) follows a central result of this work; 
i.e. the high energy nuclear fluid viscosity 
\begin{equation}
\eta =\eta_{ideal}
\left(\frac{T-T_0}{T} \right)^{9/2}
\exp\left[{\Phi \over k_B(T-T_0)}\right],
\label{VString11}
\end{equation}
with the ideal quantum viscosity determined as 
\begin{eqnarray}
\eta \approx \eta_{ideal}\ \ {\rm if}
\ \ T\gg T_0,
\nonumber \\ 
\frac{\eta_{ideal}}{2\pi \hbar}=\bar{n},
\nonumber \\ 
\bar{n}=\frac{\rho c_\infty^2}{\Phi }
\le \frac{\rho c^2}{\Phi}\ .
\label{VString12}
\end{eqnarray}
Eqs.(\ref{VString11}) and (\ref{VString12}) for the viscosity of a stringy QCD fluid 
phase is quite similar to the viscosity of a polymer chain glass. As the temperature 
is lowered from above to \begin{math} T_0 \end{math}, the viscosity becomes 
extraordinarily high and the fluid behaves more as a glass then as an ideal fluid. For 
the ultra high energy collisions of heavy nuclei, the duration of the collision is 
sufficiently short that \begin{math} T\gg T_0 \end{math}. For such a range of hot 
string temperatures, the fluid viscosity is extraordinarily low, i.e. the viscosity  
is the quantum limited value \begin{math} \eta \approx \eta_{ideal} \end{math}. The  
QCD fluid phase is thereby ideal.

\section{Conclusion}
\label{conc}

We have derived Kubo formula sum rules which allowed for the determination 
of the transverse viscosity in terms of the relaxation time of the pressure tensor. 
We then considered the quantized unit of circulation which contributes to 
turbulent eddy currents in the fluid. If the the quantum turbulent Feynman number 
\begin{math} {\cal F} \end{math} is to exceed classical turbulent Reynolds number 
\begin{math} {\cal R} \end{math}, then there exists a quantum lower bound to viscosity 
which has been previously conjectured\cite{Shuryak:2005}. Nuclear matter, as it appears 
in ultra high energy heavy nuclei scattering, was thought by some workers to have the 
formal conformal symmetry associated with deconfined glue at low baryon density. Here, 
the string fragmentation model was employed wherein the quarks and anti quarks are 
connected to one another by strings whose electric flux tubes consist of glue. The notion 
of a low baryon density nuclear fluid is thereby defined as a very stringy gluon 
liquid with not so many quarks and anti quarks. This stringy liquid is not quite the same 
as a deconfined quark gluon plasma. The viscous properties of a QCD stringy liquid is  
as described by Eqs.(\ref{VString11}) and (\ref{VString12}). The stringy liquid is 
closely analogous to a polymer glass. While the viscosity grows quite high 
as the glass (Hagedorn) temperature is approached from above, for temperatures well 
above the glass temperature \begin{math} (T\gg T_0) \end{math}, the viscosity is 
quite small and quantum limited. These small viscosity values imply and almost ideal 
ultra high energy nuclear fluid.

\end{document}